\documentclass[11pt]{article}
\usepackage{moriond,epsfig}

\def\Journal#1#2#3#4{{#1} {\bf #2}, #3 (#4)}

\def\NIMA{{\em Nucl. Instrum. Methods} A}

\def\PLB{{\em Phys. Lett.}  B}
\def\PR{\em Phys. Rep.}

\def\be{\begin{equation}}
\def\ee{\end{equation}}
\def\bea{\begin{eqnarray}}
\def\eea{\end{eqnarray}}

\begin{document}
\vspace*{4cm}
\title{DARK MATTER SEARCH AT BOULBY MINE}

\author{ R. LUSCHER }

\address{on behalf of the Boulby Dark Matter Collaboration\\
(RAL, Imperial College, Sheffield, UCLA, Texas A\&M, Pisa, ITEP, Coimbra, Temple and Occidental)\\
Rutherford Appleton Laboratory, Chilton,
Didcot OX11 0QX, United Kingdom}

\maketitle\abstracts{
The Boulby Dark Matter Collaboration (BDMC) is running a WIMP Dark Matter 
research programme in the underground laboratory of Boulby Mine. 
The two axes of the programme are based on (1) liquid Xenon (LXe) as the WIMP
target and (2) directional detection in low pressure gas detectors.
ZEPLIN-1 is a 3.1kg of LXe scintillation detector with a background
discrimination based on Pulse Shape Analysis.
A new limit from a preliminary analysis is presented.
Further generation setups with improved background discrimination tools
(as the ionisation is also recorded) are in construction. 
ZEPLIN-2 and ZEPLIN-3, are predicted to be sensitive to rate of 0.1-0.01 events/kg/day within 2 years of data, with fiducial masses of 30kg and 6kg, respectively. 
These are important steps towards the design of a tonne-scale LXe Dark Matter
detector array. 
The DRIFT programme relies on discrimination by tracking and
directionnal detection. 
This enables the measurement of any sidereal modulation, as the earth's rotation changes 
the flux angle of Galactic WIMPs. 
A low pressure 1m$^3$ chamber (DRIFT-1) is currently taking data underground. 
A setup for a bigger volume with increased spatial resolution (enabling the pressure, and hence the target mass to be increase) is in the designing process.
}

\section{Introduction}
The quest for Dark Matter is among the most fundamental of all astro-particle physics questions.
From the recent WMAP results on the cosmic microwave background\cite{map03}, we expect that 23\% of all matter of the universe is due to Cold Dark Matter, for 
which the SUSY-WIMP model provides an excellent candidate.
As new collider physics results appeared, the theoretical predictions for the rates of interaction between WIMPs and targets have gone down to the level of 10$^{-2}$ events/kg/day.
To detect WIMPs at these levels requires very sensitive and massive targets.

The UK Dark Matter Collaboration (RAL, Imperial College, University of Sheffield, University of Edinburgh) has been running a Dark Matter programme at Boulby Mine for more than a decade.
Competitive limits on WIMP cross section were set with NaI targets in 1996\cite{Smi96}.
The current programme consists of two axes: detectors based on xenon, which has a better background discrimination potential, have been developed and are currently running or being commissioned; a gaseous detection device has been constructed and is operating at Boulby with the intention of developing directional detectors.
Both projects are done in collaboration with many international groups: UCLA, Texas A\&M, Pisa, ITEP, Coimbra, Temple and Occidental.

\section{ZEPLIN: liquid Xenon as Dark Matter target}
The ZEPLIN project (Zoned Proportional scintillation in LIquid Noble gases), 
takes advantage of the particularly appropriate properties of Xe:
heavy nuclei for a large spin-independent coupling and appreciable 
abundance of isotopes with spin for a large spin-dependent coupling.
Low background Xe is available.
LXe is also known as a good scintillator: emitting in the UV region (175~nm), 
it enables a low energy threshold.
Moreover, the interaction process in LXe owns characteristics
which translate into a potentially high background discrimination.

Any recoil in LXe give rise to both ionisation and excitation of Xe atoms.
The excitation result in the emission of a 175~nm photon from either a 
singlet (with decay time $\sim3$ ns) or a triplet state ($\sim27$ ns).
The ratio single/triplet is 10 times bigger for nuclear recoil compared to 
electron recoils.
In the absence of an electric field, the ionisation recombines to produce 
further excited Xe atoms. The recombination time depends on the ionization density: for nuclear recoil, it is very high and recombination very fast.
For electron recoil, the lower density leads to longer times.

\subsection{ZEPLIN-1, a single phase detector}
\begin{figure}[htb]
\centerline{\epsfig{file=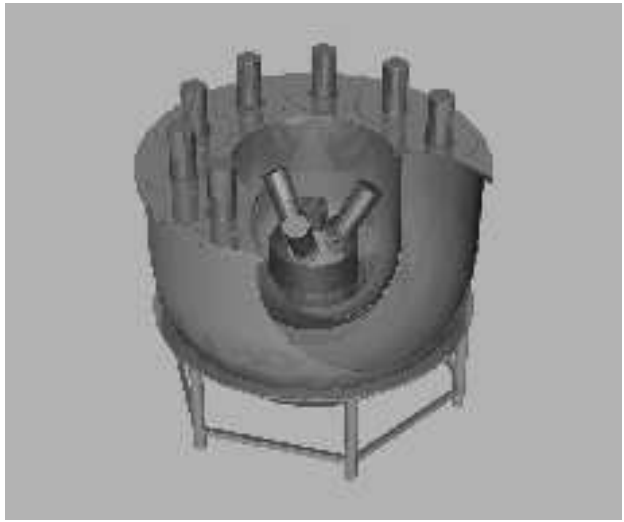,width=8.2cm}}
\caption{The ZEPLIN I design: a pure LXe scintillator with 4~kg active mass.}
\label{ZepI}
\end{figure}

The first detector of the ZEPLIN array is based on a pure scintillator design. 
The LXe is viewed by 3 PMTs through silica windows (figure \ref{ZepI}).
Between the fiducial volume, which contains 3.1~kg, and each 
window, there is a volume of LXe which is optically isolated (ie. in 
which most of the signals appear only in the corresponding PMT), thus 
acting as self-shielding.

The detector is enclosed in a 1 tonne active Compton veto shielding, based 
on PXE liquid scintillator and viewed by 10 x 8" hemispherical PMTs.
Its function is to veto gamma activities from the PMTs and from the 
surroundings.
From Monte-Carlo simulations, a veto efficiency of 80-90\% below 100~keV 
has been estimated.

The detector is triggered by a 3-fold coincidence of a single p.e. in each tube.
With a light yield of at least 1.5 p.e./keV in the data runs, it gives a 2~keV threshold.
The trigger efficiency has been calculated using Poissonian statistics
The signals are digitized using an Acqiris cPCI based DAQ system.
The dead-time is smaller than 2~$\mu$s, its efficiency hence bigger than 99.9\% during normal data runs.

\begin{figure}[htb]
\vspace{9pt}
\centerline{\epsfig{file=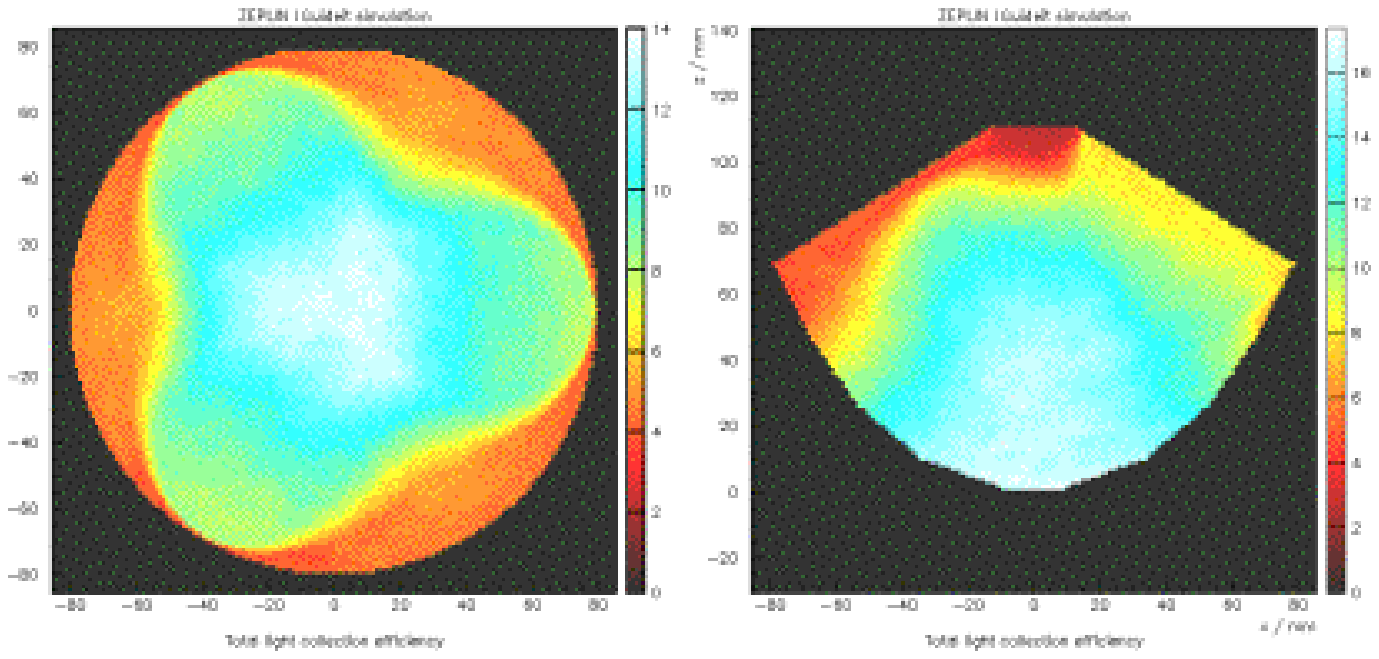,width=9cm}}
\centerline{\epsfig{file=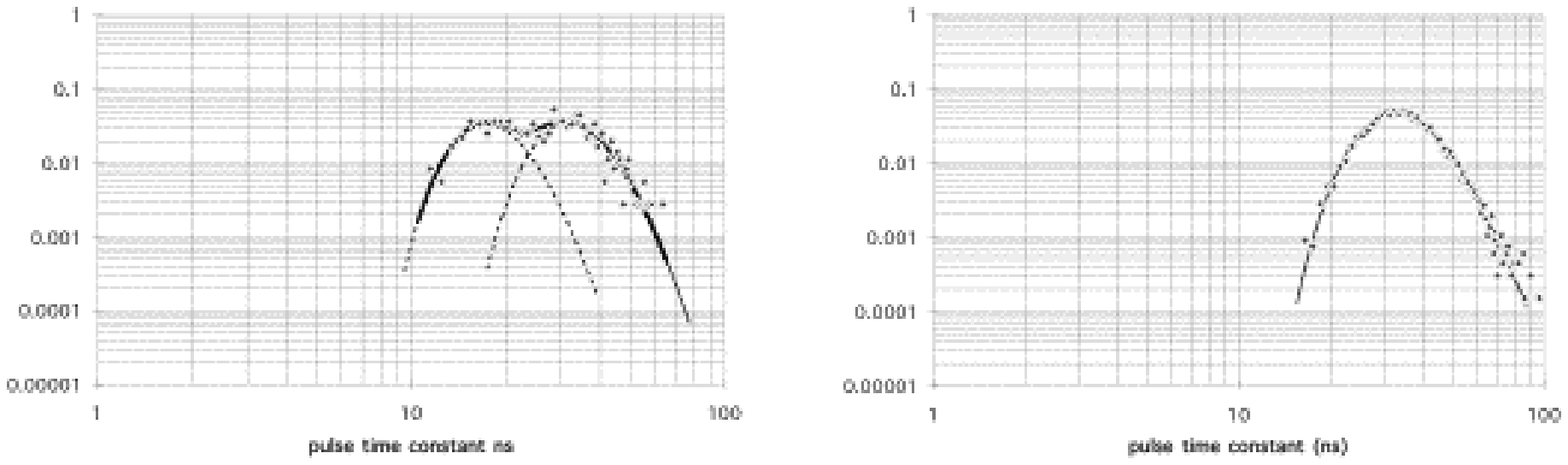,width=10cm}}
\caption{Light collection efficiency of ZEPLIN-1 (top): scan through detector girth (left) and through one PMT plane (right). Scintillation time constant distributions (bottom): an Am/Be neutron source calibration (left) and a comparision between data a gamma calibration (right) are shown. Superimposed is a $\Gamma$ density distribution in 1/t.}
\label{lightcoll}
\label{distri}
\end{figure}

Daily energy calibration is performed with a $^{57}$Co source automatically placed between target and veto.
The 122~keV $\gamma$s convert within the bottom ~3~mm in the target, making it a calibration point source.
A 30~keV K-shell X-ray is also observed in the spectrum, its presence has been confirmed through a Geant4 simulation.
A full light collection simulation has been performed (figure \ref{lightcoll}), showing variations of efficiencies from a maximum of 18\% at the bottom of the target down to 4\%  just below the Xenon delivery line.
This affects the measured energy of an event and has been observed in higher energy gamma calibrations ($^{60}$Co, $^{137}$Cs sources): as different fractions of the target are illuminated, the peak position reflects the reduction in light yield. 
The observations match well the light collection efficiency simulation.

The light collection study has been extended to each tube separately. 
This permits to assess the capability of rejecting turrets events.
A turret parameter (S$_3$) translates the asymmetry of events into a number. 
S$_3$ = 0.81 indicates an event with all the signal in a single PMT, while a completely symmetric event would give a null value.
In a turret event, in the ideal case of no loss of light within the chamber, one single PMT is expected to get 66\% of the light, resulting in S$_3$ being 0.41.

Purification of the Xe gas is performed by using an Oxisorb, as well as by pumping on the frozen Xe and subsequent fractionation of the Xe gas.
From the observed background spectrum, we can extract a $^{85}$Kr contamination level of 10$^{-17}$ atom-per-atom in the LXe, assuming it is the main source.
A source for low Kr contamination (40~ppb, while research grade specifies 5~ppm) Xenon has been found.
With this new Xenon, a background reduction of a factor 3 has been achieved.
Moreover, the scintillation performance has also increased, as a light yield of 2.5 p.e./keV has been observed.
This is probably due to the purification used to remove the Kr, which also removes other contaminants (such as CO$_2$).

Background discrimination is provided by the difference in time constant of the scintillation pulse induced by nuclear or electron recoils (figure \ref{distri}).
Neutron and gamma source calibrations done on the surface have shown that a typical nuclear recoil time constant is 55\% of a typical electron time constant, currently measured down to 4~keV.

\begin{figure}[htb]
\centerline{\epsfig{file=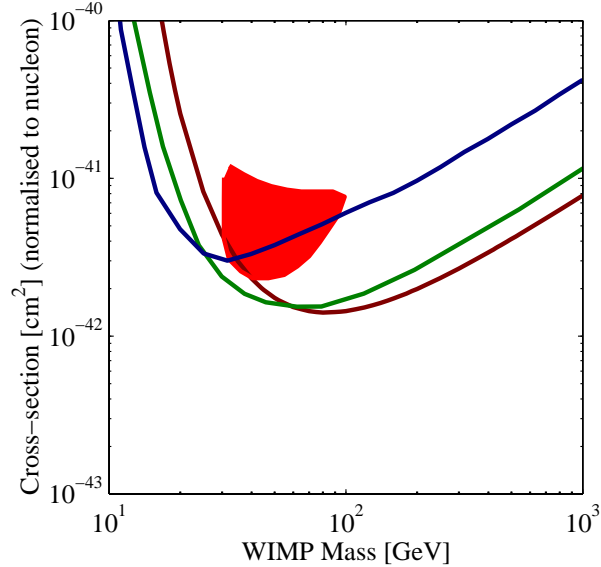,width=8cm}}
\caption{Preliminary ZEPLIN-1 limit on WIMP-nucleon cross section from 75 days of data, in comparison with CDMS (left curve), Edelweiss (left curve) and the DAMA result (surface)}
\label{limit}
\end{figure}

The 90\% C.L. limit on nuclear recoil is extracted by studying the monotonically rising edge of the time constant distributions and comparing it with calibration data from gamma sources or known backgrounds.
The events tagged by the Compton veto (mainly high energy gammas from the PMTs) form the latter.
The upper limit on the number of nuclear recoils in each energy bin is then used to calculate the WIMP-nucleon cross section.
The limit extracted from 230 kg$\times$days of data in a preliminary analysis is shown in figure \ref{limit}.
The data taken with the low Kr contamination Xenon (50.2 kg$\times$days) are currently being analyzed. 

\subsection{ZEPLIN-2, a double phase detector}
The two-phase detector measures both scintillation and ionization produced by interacting particles separately.  
It comprises a liquid target, gaseous xenon layer and an applied electric field.  
Any interacting particle produces excitation and ionisation within the liquid 
xenon target.  
An applied electric field partially suppresses recombination, drifting electrons upwards through the liquid to the gaseous phase.
There, a wire planes define a high field region in which avalanche occurs and 
electroluminescence is created.
The latter can then be recorded by the same PMTs which have already read the primary scintillation.
The proportion of ionisation and excitation released depends on the $dE/dx$ 
of the particle interaction.
Electron recoil produces more ionisation, while for nuclear recoils, the 
excitation is more important.
This technology has been tested in a 1~kg prototype\cite{Wang98,Wang00} and 
shown very promising discrimination power. 

\begin{figure}[htb]
\centerline{\epsfig{file=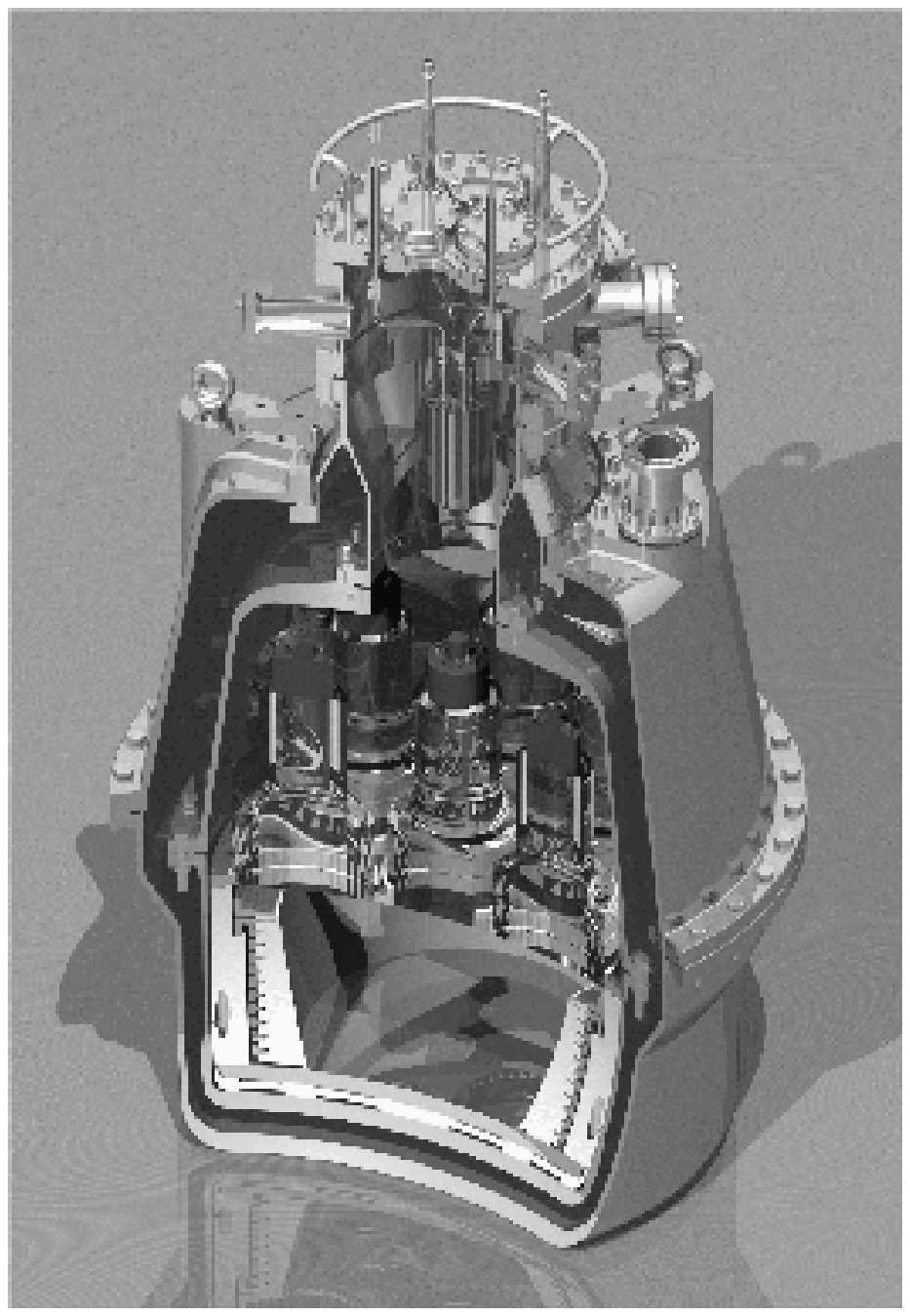,width=5.3cm} \hspace*{1cm}
\epsfig{file=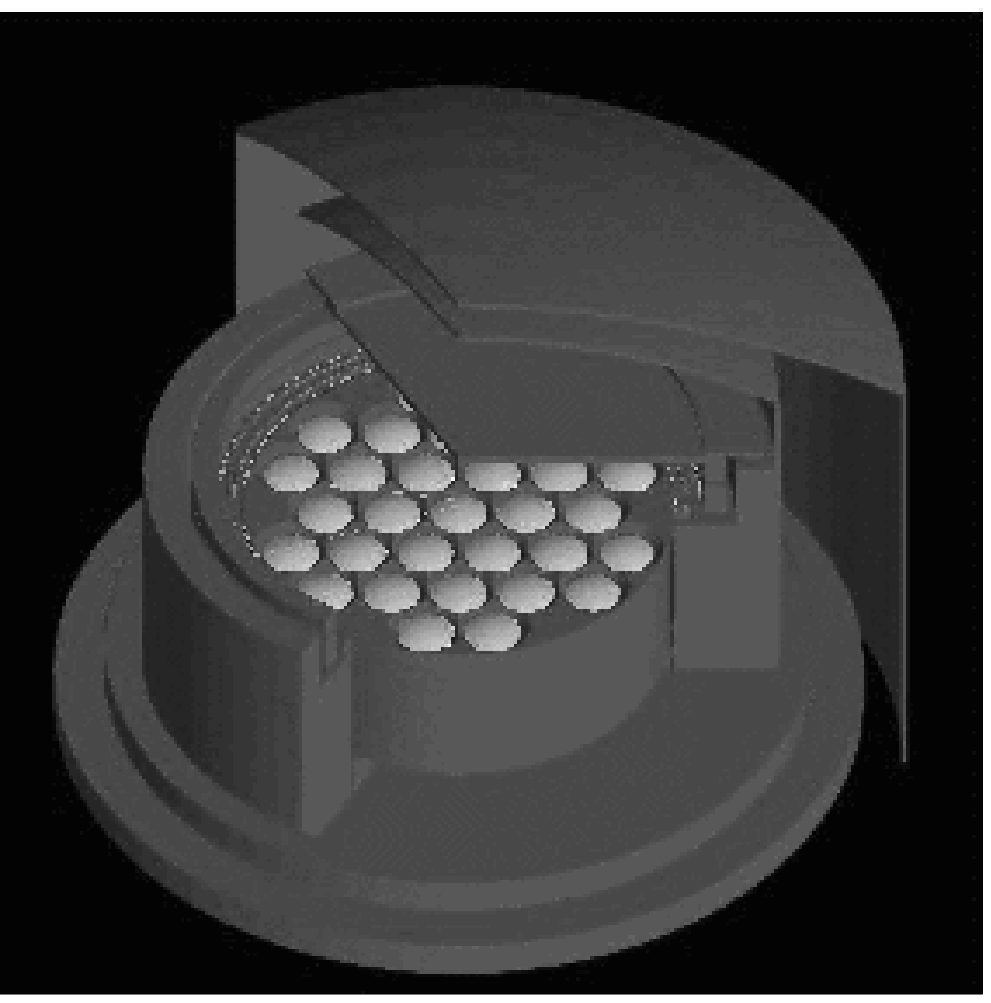,width=7.cm}}
\caption{The ZEPLIN-2 (left) and ZEPLIN-3 (right) double-phase designs.}
\label{zep23}
\end{figure}

The ZEPLIN-2 design is based on this technology.
The detector contains a target mass of about 30~kg (figure \ref{zep23}). 
It will be surrounded by the same liquid scintillator Compton veto as ZEPLIN I.
Sensitivity to rates of about 0.1-0.01~events/kg/day can be reached 
with 2 years of data.
The construction is well underway and commissioning has started.
The installation in the Boulby Mine laboratory is scheduled for early 2004.

\subsection{ZEPLIN-3: dual phase with high electric field}
The ZEPLIN-3 experiment (figure \ref{zep23}) operates with a high electric field (~8 kV/cm in the liquid target) to observe scintillation and  ionisation from nuclear recoil events.
The ratio of the the two signals gives discrimination between nuclear recoils and background gamma rays.
The active liquid xenon volume is cylindrical with a depth of 3.5 cm and radius of 2 cm.  
Both VUV signals are observed by an array of 31 photomultiplier tubes submerged in the liquid.  
The larger electroluminescence signal is used to determine the position of events, allowing definition of a fiducial volume.

This design improves the background discrimination and allows a lower threshold.
The fiducial volume is constrained by the limits of High Voltage: with an active liquid depth of 3.5~cm, the design has a rather small fiducial mass of 6~kg.
Nevertheless, similar sensitivities as ZEPLIN-2 will be reached, as the discrimination 
and the threshold are improved.
More details can be found in \cite{Sumn02}.
Schedule for installation in Boulby Mine is mid-2004.

\subsection{Towards 1 ton}
The prospective limits of ZEPLIN-2 and ZEPLIN-3 in terms of WIMP-nucleon cross section lies somewhere in the regions of 10$-8$ pb.
Many SUSY models predict lower cross sections, down to 10$-{10}$ pb regions.
To cover these predictions, a detector mass of 1 tonne might well be needed.
We are currently studying different design possibilities for a scale-up towards 250~kg modules, with the constraint of achieving a low energy threshold.

\section{DRIFT: a Directional Detector}
A definitive WIMP signature can be searched for by measuring sidereal variations of nuclear recoil directions. 
A detector sited at the Boulby Mine will see the mean recoil direction rotate from downwards to southwards and back again over one sidereal day.
It opens also the prospects for directly probing galactic halo structure and dynamics.

\begin{figure}[htb]
\vspace{-0.5cm}
\centerline{\epsfig{file=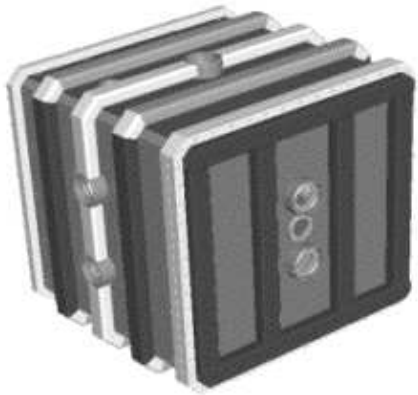,width=6.5cm}}
\caption{DRIFT-1: a 1 m$^3$ Time Projection Chamber.}
\label{drift}
\end{figure}

DRIFT-1, currently in operation at Boulby mine, consists of a 1~m$^3$ Time Projection Chamber of low pressure (about 40 Torr) CS$_2$ gas, with a drift field of 260V/cm (figure \ref{drift}).
The free electrons produced in an ionization track combine with the electro-negative CS$_2$; these negative ions are then drifted towards a Multi-Wire Proportional Chamber readout.
In such a low pressure, electron drift would show a very large track diffusion; negative ion drift, on the other hand, has been shown to enable sub mm resolution with 0.5~m drift length\cite{Mart00}.

The tracking has not only a directional issue: comparison of track length to deposited energy ($dE/dx$) is a powerful background rejection tool, as electron recoil and alpha tracks extend significantly more than nuclear recoils.
The efficiency of discrimination is such that no passive $\gamma$ shielding is needed.

DRIFT-1 has been underground since mid-2001.
It is mainly intended as a proof of feasibility, as an assessment of the sensitivity to tracking, background rejection and directionality.
Calibration runs with a neutron source ($^{252}$Cf) placed on top or on the side of the detector have shown differences in the angular distributions of the nuclear recoil.
As such, DRIFT-1 has fulfilled its purpose as prototype.
It is now intended to run it without any shielding in order to assess the neutron background in our lab.
After that, neutron shielding will be installed around the detector.

The tracking resolution of DRIFT-1 is limited by the pitch of the MWPC plan (1~mm).
An R\&D programme is currently testing the possibility to use alternative read-out techniques with narrower pitch, such as MICROMEGAS.
This would allow to increase the pressure in DRIFT, hence the target mass, without loosing any sensitivity.
Also alternatives to CS$_2$ as the target gas are studied.
This programme will lead to the design of a new generation detector, DRIFT-2.

\section*{Acknowledgments}
The BDMC would like to thank Cleveland Potash Ltd for the continued support at Boulby mine.

\end{document}